\documentclass{article}
\usepackage{graphicx} 
\usepackage{indentfirst}

\usepackage{amsmath}
\usepackage[final]{microtype}
\usepackage{xcolor}
\usepackage[hidelinks]{hyperref}
 \hypersetup{ 
     colorlinks=true, 
     linkcolor=blue, 
     filecolor=blue, 
     citecolor = black,       
     urlcolor=cyan, 
     } 
\usepackage{subfigure}

\title{Median-of-Means Sampling for the Keister Function}
\date{April 30, 2024}
\author{Bocheng Zhang}

\begin{document}

\maketitle

\section{Introduction}
Integration, a cornerstone of scientific advancement since its discovery in the late 17th century, has evolved significantly with scientific and technological progress. Integration plays a fundamental role in physics, chemistry, finance, and various other disciplines involving mathematics. Computing the integral, though, has been a challenging problem: even as early as the 17th century, mathematicians discovered integrals that could not be computed analytically \cite{Goldstine1977}. This gap gave rise to the study of numerical integration, which utilizes numerical methods to calculate approximate answers of integrals. 

Studies in numerical integration have traditionally focused on theoretical approaches, such as deriving approximation formulas for various integrals of interest. However, in the 20th century,  numerical integration research began to rely more on the computational advantages provided by computers. The 1950s witnessed a surge in numerical integration research driven by advancements in computer science. Among various methods developed during this period, the Monte Carlo (MC) method, which utilizes statistical sampling for integral calculations, emerged as particularly notable \cite{Metropolis1987}. 

The Monte Carlo method holds its roots from Los Alamos during the Manhattan Project. Proposed by physicists Stan Ulam, John von Neumann, and Nicholas Metropolis, Monte Carlo methods simulate the process of computing an integral by sampling random points in a given space (typically an $n$-dimensional unit cube) and calculate the average of these points to approximate the integral \cite{MC1949}. The method became successful in solving problems in particle, high energy, and--of course--nuclear physics and inspired the development of algorithms for computational physics and applied mathematics. 

Fueled by increasing computational power and data volume, demand for numerical integration saw a burgeoning in the late 20th century. Many fields ranging from molecular biology to finance rely on mathematical models that require computer-aided integration. Consequently, researchers developed Quasi-Monte Carlo (QMC) methods, a variant of MC, tailored for high-dimensional problems and intensive computing needs. QMC outperforms MC in high-dimension integration, offering greater precision and efficiency \cite{Caflisch_1998}. QMC methods are widely used in physics and finance. 

More recently, the field of numerical integration has been enriched by Randomized Quasi-Monte Carlo (RQMC) methods. While traditional MC methods occasionally utilize randomness, RQMC utilizes repeated simulations to improve convergence rates and reduce the sample size needed for accurate results. Despite slightly limited theoretical backing, RQMC has shown success in numerous applications, suggesting its potential effectiveness and efficiency. 

\section{Literature Review}

A significant advantage of RQMC over traditional Monte-Carlo is its faster error convergence rate \cite{Lecuyer2002}. In other words, RQMC is capable of achieving results that are more precise given the same computation resources, or achieve a same level of precision given less resources compared with traditional MC. A factor that plays into the convergence rate and performance of these integration methods is the function that is being integrated. Depending on specific parameters of the function, such as the smoothness of the function, integration performance could vary greatly \cite{GodaLEcuyer}. Such variability is largely due to how the integral is computed: traditional QMC algorithm computes several different approximations of the integral and takes the arithmetic mean of these integrals as the final result \cite{Glasserman2003,Niederreiter1992}. This is known as the mean-of-means stopping criterion, since it takes the sample mean of integral approximations, which is the sample mean of the integral at different points. The mean-of-means sampling scheme is known for consistent performance in computation speed and theoretically predictable convergence rates, and therefore has been used as the standard algorithm for performing RQMC integration. 

This paradigm, however, is being changed by the recently-developed median-of-means sampling method, jointly proposed by Art Owen and Zexin Pan from Stanford University \cite{OwenPan1dmom,multidMomOwenPan}, Pierre L'Ecuyer from the University of Montreal, and Takashi Goda from the university of Tokyo \cite{GodaLEcuyer}. This approach, which differs from traditional sampling methods, involves evaluating the functional values of integrands at various points within an n-dimensional space, repeating the process multiple times, and determining the final result by computing the median of all calculated outcomes \cite{OwenPan1dmom,multidMomOwenPan}. This method has been promising in efficiently handling totally-random point sets.

\begin{figure} [h]
    \centering
    \includegraphics{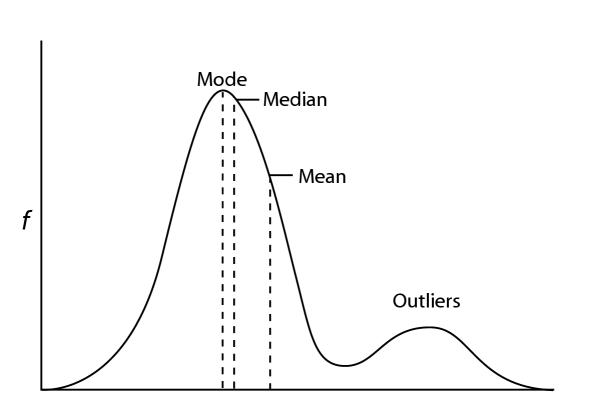}
    \caption{The effects of outliers the mean and median of a nearly normal distribution}
    \label{fig:mean-median}
\end{figure}

The idea behind median-of-means is simple: RQMC methods approximate integrals by sampling points and taking their averages. The points, however, are randomized, so the results may vary. To reduce variability, RQMC methods repeat this process and compute, for example, 20 different averages from 20 point batches. From here, the procedure begins to differ: while the traditional mean-of-mean take the mean of this sample, median-of-means takes the median \cite{GodaLEcuyer}. Why then, is the median-of-means more reliable than its counterpart? Intuitively, it is because the median of a distribution is less likely to be influenced by an outlier. 

As shown in the above figure, for a distribution that is approximately normal, outliers would have a greater skewing impact on the mean over the median. The idea is similar for QMC integration, as most of the sampled integrals follow a normal distribution, and the actual value tends to exist around the center of such distribution \cite{OwenPan1dmom,multidMomOwenPan}. Therefore, it is natural to take the median of these integrals instead of the mean in order to better approximate the actual value. 

Beginning in 2021, Owen and Pan began to analyze the median-of-means sampling method based on digital nets, a type of point distribution \\\cite{OwenPan1dmom}; in parallel, L'Ecuyer and Goda started to study the properties of the method based on lattices, another type of point distribution \cite{GodaLEcuyer}. Both groups were quick to obtain fruitful results: they were able to demonstrate the superiority of this sampling method for certain types of functions under certain constraints, but faced challenges in showing the robustness of the algorithm under more general circumstances. 

L'Ecuyer, Goda, Owen, and Pan laid the foundation for the new paradigm of median-of-means, but much work is to be done. In 2023, these researches and many others made theoretical advancements on the properties of median-of-means \cite{10.5555/3643142.3643179}, \cite{owen2023gain}, \cite{gnewuch2023computable}. However, they have emphasized the need for conducting numerical experiments to further test the median-of-means sampling method. 

As both a highly theoretical and practical study, RQMC benefits equally from theoretical predictions and numerical trials \cite{Niederreiter1992, Glasserman2003}. Empirical experiments can demonstrate the effectiveness of certain algorithms in addressing practical problems, such as measuring options in finance\\ \cite{AsianOption}, and expose certain limitations to the applicability and robustness thereof. 

This study attempts to address the gap in understanding of median-of-means sampling method by conducting numerical experiments on the Keister function--widely used in nuclear physics problems, and a common test function for QMC \cite{Keister1996}. This function has not been previously tested by studies in median-of-means, but has been used as a test function in previous studies \cite{Papageorgiou_1997,doi:10.1137/S1064827599356638}. The Keister function is used as a primary testing function because some of its values are calculable for certain dimensions, thus allowing a direct comparison for the approximation effect by computing differences between the approximated result and the actual answer. Furthermore, no studies have tested the Keister function on the median-of-means sampling method at the time of this study, accentuating the need for applying the method on the Keister.

Given the same sample size, dimension, and point distribution, this study hypothesizes that the median-of-means sampling method outperforms the mean-of-means sampling method by achieving a smaller error when integrating the Keister function.

\section{Methods}

\subsection{Overview}

 The Keister function will be tested over two types of point distributions--lattices and digital nets--that were used in Owen and Pan, Goda and L'Ecuyer's studies \cite{GodaLEcuyer,OwenPan1dmom}. The experiment contains a control group and a treatment. The control group being the traditional mean-of-mean sampling method, and the treatment, of course, being the median-of-means method. The experiment will compare the error convergence rate--the speed at which the difference between the approximated value and the actual value decreases. The results will then be analyzed on a comparison graph and using statistical methods and can be found in \href{https://github.com/bdavidzhang/research/blob/main/median_of_means.ipynb}{this notebook}.

\subsection{Lattice and Digital Net}

\subsubsection{Lattices}
This section provides an overview of the lattice and digital nets. lattices are mathematical structures used to generate low-discrepancy sequences of points, which are employed for numerical integration and approximation of high-dimensional integrals.

Lattices are used in QMC methods to generate sequences of points that are more uniformly distributed in the n-dimensional unit cube than pseudo-random sequences. This property is achieved by carefully constructing the basis vectors of the lattice to ensure that the points are well-spaced and do not cluster in certain regions \cite{Glasserman2003}.

\begin{figure} [htbp]
    \centering
    \includegraphics[width=0.45\linewidth]{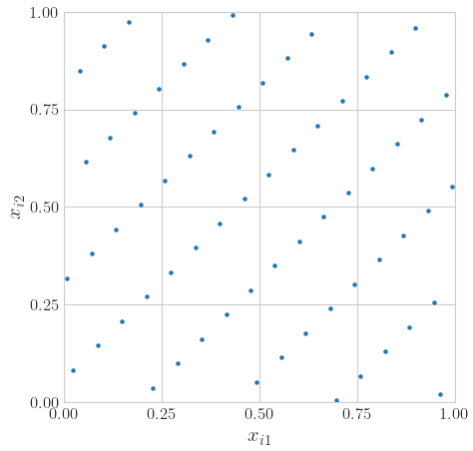}
    \includegraphics[width=0.45\linewidth]{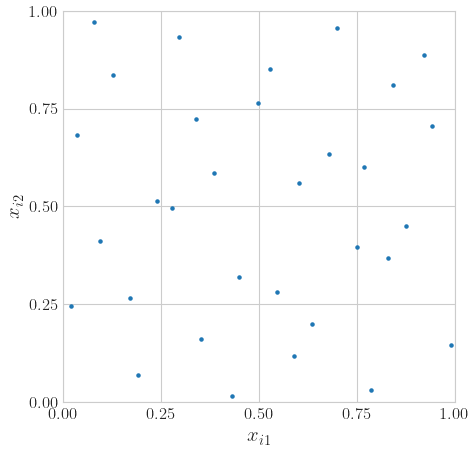}
    \caption{Left: a 2-dimensional lattice, right: a 2-dimensional digital net. Both point sets contain 64 points and are generated by QMCPy.}
    \label{fig:latdnet}
\end{figure}

Lattices are extensively studied and used in QMC methods because they can provide significantly faster convergence rates for high-dimensional integrals compared to classical Monte Carlo methods, which rely on pseudo-random sequences. However, constructing efficient lattices and analyzing their properties can be computationally intensive, especially in higher dimensions. In other words, a good lattice is hard to find. 

L'Ecuyber and Goda demonstrated the effectiveness of median-of-means for lattices that are totally randomized. Traditionally, high dimensional lattices are constructed by carefully searching for lattices with high efficiency \cite{Glasserman2003}. This is because the mean-of-means sampling method is sensitive to outliers, and poorly-constructed lattices may produce many outliers in computation. The median-of-means method, however, is more resilient to these disturbances\cite{GodaLEcuyer}.

\subsubsection{Digital Nets}

Digital nets are a type of low-discrepancy sequences used in QMC methods for numerical integration and approximation of high-dimensional integrals. They are constructed using a digital construction method, which is based on the digital expansion of fractional numbers.

In the digital construction, each point in the n-dimensional unit cube is represented by a sequence of digits in base $b$, where $b$ is typically chosen to be a prime number \cite{Glasserman2003} (in QMCPy, $b=2$). The digits are arranged in a specific way to ensure that the resulting sequence of points has a low discrepancy, meaning they are well-distributed and do not cluster in certain regions.

Digital nets are widely used in QMC methods due to their strong theoretical properties and practical effectiveness in high-dimensional numerical integration problems. However, the construction and analysis of digital nets can be computationally intensive, especially in higher dimensions, and the choice of generator matrices plays a crucial role in the quality of the resulting sequence. Similar to lattices, digital nets also rely on high quality generating matrices, which are hard to generate. 

\subsection{QMCPy}

QMCPy is a python library developed by researchers Sou-Cheng Choi, Fred Hickernell, R. Jagadeeswaran and Aleksei Sorokin from the Illinois Institute of Technology \cite{choi2021quasimonte}. It is designed to provide tools for generating low-discrepancy sequences, such as lattices and digital nets, which are commonly used in QMC methods for numerical integration and approximation.

QMCPy is selected for this experiment for several reasons: first, it is open-source and widely used by researchers in QMC such as Owen and Pan \cite{multidMomOwenPan}. Second, the library is well-documented and includes examples to facilitate usage \cite{Sorokin2022}. Third, the library provides mechanisms for ensuring reproducible results. All the computed results for this experiment are seeded and can be conditionally reproduced. 

\subsection{Statistical Analysis}

Statistical analysis for this experiment is carried out via a comparison graph, which is a typical tool for QMC researchers, featured in \cite{GodaLEcuyer, OwenPan1dmom, owen2023gain, doi:10.1137/S1064827599356638}, and many other papers. This study imitates the design of the comparison graph in \cite{GodaLEcuyer}--a graph on the relationship between sample size and error. 

To reflect the changes more accurately, graphs plotted on a logarithmic scale, with the sample size (independent variable) increasing exponentially and error (dependent variable) decreasing exponentially. Two lines will be plotted, a blue line corresponding to the median-of-means sampling method, and an orange line corresponding to the mean-of-means. The differences between the two graphs are marked in red. To better illustrate the differences in performance for the sampling methods, graphs of the difference between the error in the median-of-means and the mean-of-means is provided. This graph is inspired by \cite{doi:10.1137/S1064827599356638}. 

A comparison graph will be generated for each dimension, and each type of point distribution. Four plots are generated for lattices and four for digital nets, each with a corresponding difference plot.

\subsection{Implementation}

The design for this experiment is similar to the numerical experiment performed in \cite{GodaLEcuyer}. The parameters involved in the process are: 
\begin{enumerate}
    \item $D$: dimension of the Keister integral (2,3,5,8).
    \item $P$: type of point distribution, either digital net or lattice. 
    \item $N$: sample size of the point set selected (ranging from $2^8$, $2^9$ to $2^{19}$, a total of 12 different sample sizes).
    \item $R$: number of approximations generated from randomized point sets ($R$ is set to 11).
    \item $T$: \footnote{Parameters $D,P,N,R,T$--dimension, point distribution, sample size, number of approximations, and number of trials ran--were adjusted due to limitations in computing power and QMCPy. See section 5.2 for limitations.}number of trials ran. The final result is the average of these trials ($T$ is set to 25).
    \item $I$: each approximation generated from a point set using QMC methods. $I$ is the average of all the points in a point set.
    \item $A$: the median of $R$ approximations generated in a single trial.
    \item $A'$: the mean of $R$ approximations generated in a single trial.
    \item $F$: the averaged result according to the median-of-means sampling method, the mean of all $A$s generated in $T$ trials.
    \item $F'$: the averaged result according to the mean-of-means sampling method, the mean of all $A'$s generated in $T$ trials.
    \item $S$: the solution to the Keister integral at a specific dimension. 
    \item $E$: the error according to median-of-means, $E = S - F$.
    \item $E'$: the error according to mean-of-means, $E' = S - F'$.
\end{enumerate}

For each sample size $N$ between $2^6$ and $2^{17}$, dimension $D$ belonging to either $2,3,5,8$, a  set of $N$ points are generated, denoted by $P$. $P$ is either a set of lattice points or digital net points. Then $R=11$ approximations are computed, and denoted by $(I_1,I_2,\dots,I_R)$, then the median of these approximations,
\begin{equation*}
    A = \mathrm{median}(I_1,I_2,\dots,I_R)
\end{equation*}
 is the result for this trial. The process is then repeated $T=25$ times, resulting in a collection of results $(A_1,A_2,\dots,A_T)$. The final averaged result is the mean of these $T$ results, that is, 
\begin{equation*}
    F = \mathrm{mean}(A_1,A_2,\dots,A_T).
\end{equation*}

To compute the error for this trial, 
\begin{equation*}
    E = S - F.
\end{equation*}

The above process produces the final averaged result under the median-of-means paradigm (treatment group) for a sample size $N$ and dimension $D$. The process for computing the final averaged result under the mean-of-means is highly similar. The only differences are that 
\begin{align*}
    &A' = \mathrm{mean}(I_1,I_2,\dots,I_R)\\
    &F' = \mathrm{mean}(A'_1,A'_2,\dots,A'_T)\\
    &E' = S - F'.
\end{align*}

To compare the results for a fixed dimension and a type of point distribution, the error for each sample size is collected into a tuple $(E_1,E_2,\dots,E_{12})$, in this case, $E_1$ is the error for median-of-means when the sample size is $N = 2^8$; similarly, the error for mean-of-means is collected into a tuple $(E'_1,E'_2,\dots,E'_{12})$. These errors are then plotted onto a comparison graph with the independent variable being sample size, and the dependent variable being error. Two curves are then plotted, one corresponding to the treatment curve (median-of-means) and the other the control curve (mean-of-means). To better visualize differences, both the independent and dependent variables will be plotted on a logarithmic scale.

\section{Results}

\subsection{Results for Digital Nets}

\begin{figure}[h!]
    \centering
    \includegraphics[width=0.45\linewidth]{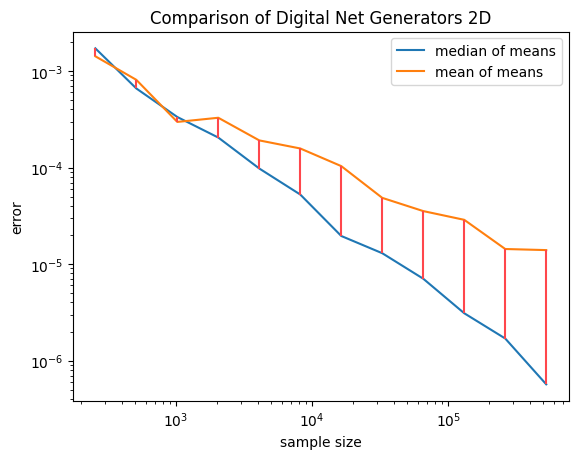}
    \includegraphics[width=0.45\linewidth]{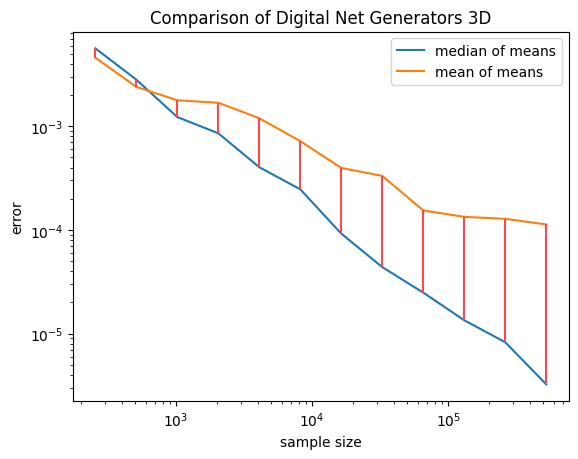}
    \includegraphics[width=0.45\linewidth]{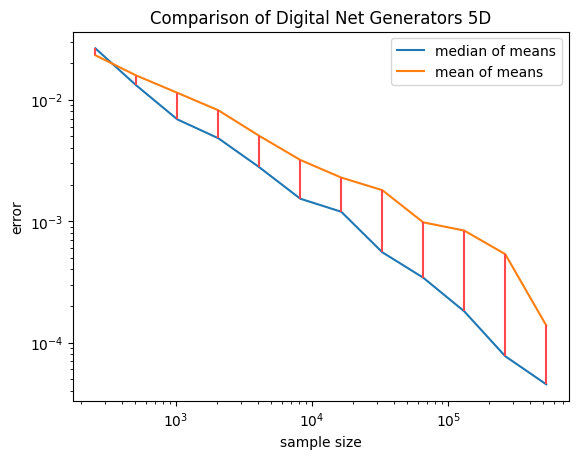}
    \includegraphics[width=0.45\linewidth]{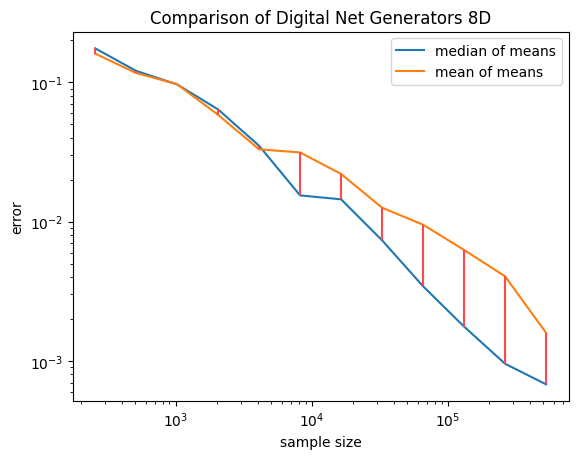}
    \caption{Comparison of digital net generators in dimensions 2,3,5,8. Results can be conditionally reproduced.}
    \label{fig:dnetcomp}
\end{figure}

\begin{figure}[htbp]
    \centering
    \includegraphics[width=0.45\linewidth]{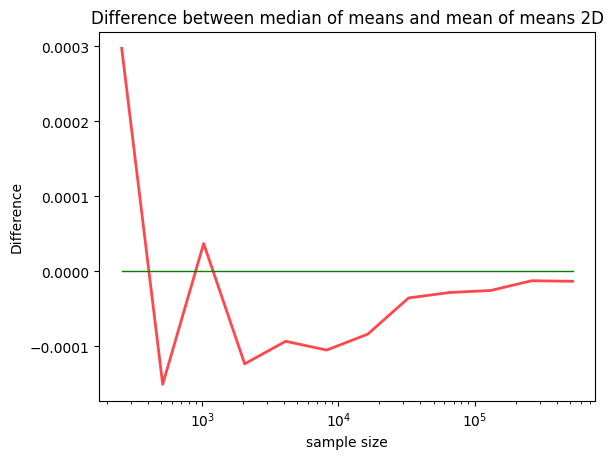}
    \includegraphics[width=0.45\linewidth]{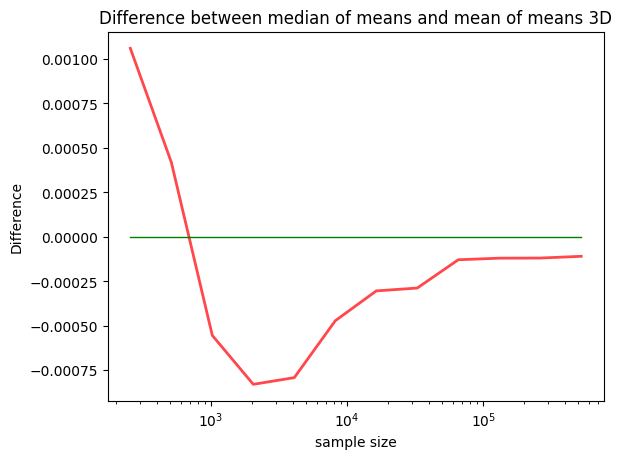}
    \includegraphics[width=0.45\linewidth]{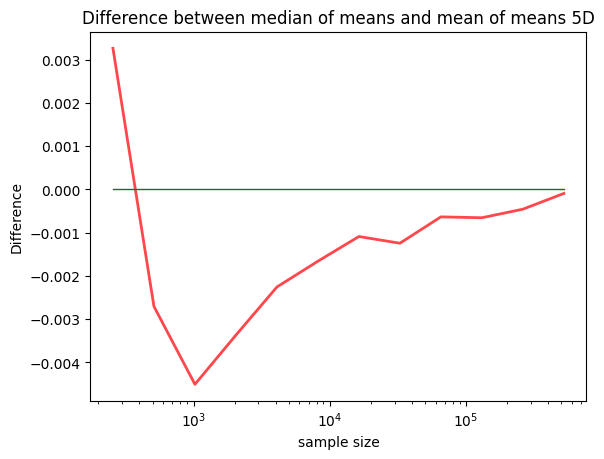}
    \includegraphics[width=0.45\linewidth]{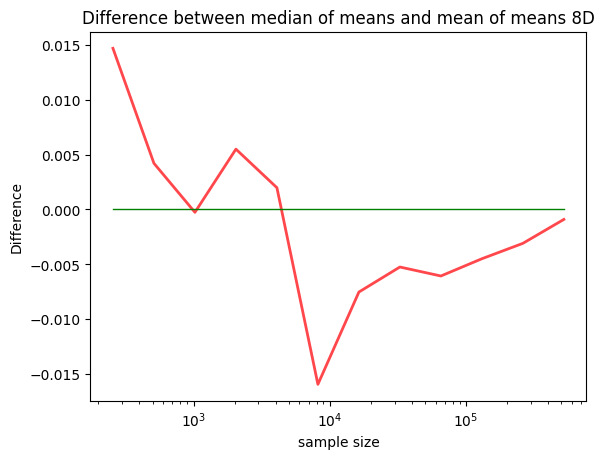}
    \caption{Difference graph of digital net generators in dimensions 2,3,5,8. Results can be conditionally reproduced.}
    \label{fig:dnetdiff}
\end{figure}

Figure \ref{fig:dnetcomp} shows the comparison between the error produced via the median-of-means and mean-of-means sampling method using digital net generators on a comparison graph. The median-of-means sampling method is colored in blue, and the mean-of-means method in orange (also reflected on the label). The vertical red lines reflect the differences between errors at a given sample size. Dimensions 2,3,5,8 were featured in the process. The graph itself is piecewise-linear (lines connecting 12 data points from 12 sample sizes ranging from $2^8$ to $2^{19}$). 

A fundamental trend that is depicted by these comparison graphs is that, the gap between the two graphs is widening as the sample size increases (the red line is increasing in length). This, however, is not suggesting that the difference in error produced by the two sampling methods is increasing. In fact, quite the opposite--the gap between the two sampling methods is decreasing and converging to zero, as shown in figure \ref{fig:dnetdiff}. The gap is widening because the differences are reflected on a logarithmic scale, hence the error is decreasing on an exponential scale, making the difference more apparent.

An overall trend for all dimensions tested is that the mean-of-means sampling method tends to be more accurate for smaller sample sizes (less than $10^3 = 1000$, as depicted in figure \ref{fig:dnetdiff}), this echos the prediction according to \cite{OwenPan1dmom,multidMomOwenPan}, that median-of-means only achieves an advantage over mean of means when the sample size is sufficiently large. 

For higher dimensions (5D, 8D), it is also noteworthy that the accuracy is quite low--around $10^{-3} = 0.001$ at best, whereas the accuracy for lower dimensions (2D, 3D) can reach around $10^{-6}$. This is largely because high-dimension integration requires a larger sample size to reach a higher accuracy \cite{Caflisch_1998}, for both lattices and digital nets.

\subsection{Results for Lattices}

\begin{figure}[h!]
    \centering
    \includegraphics[width=0.45\linewidth]{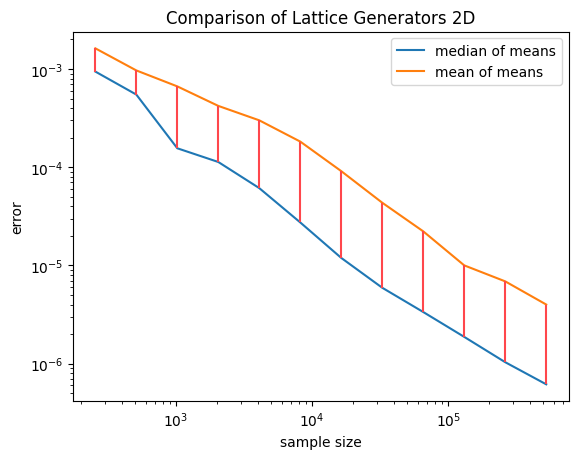}
    \includegraphics[width=0.45\linewidth]{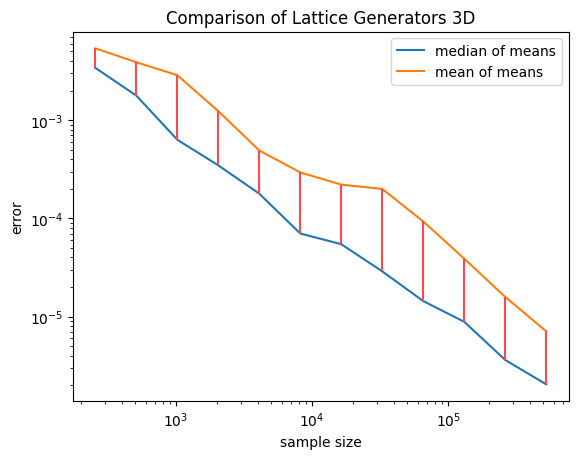}
    \includegraphics[width=0.45\linewidth]{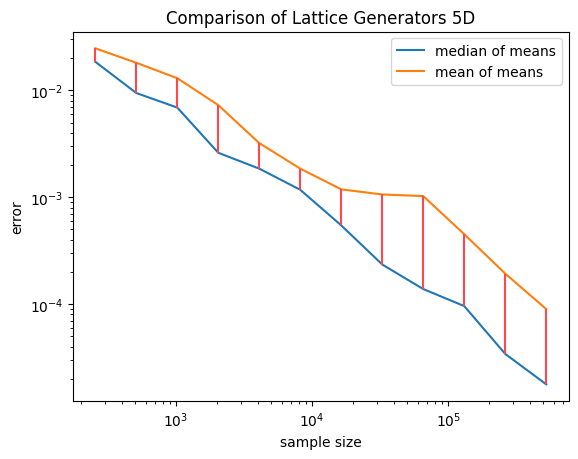}
    \includegraphics[width=0.45\linewidth]{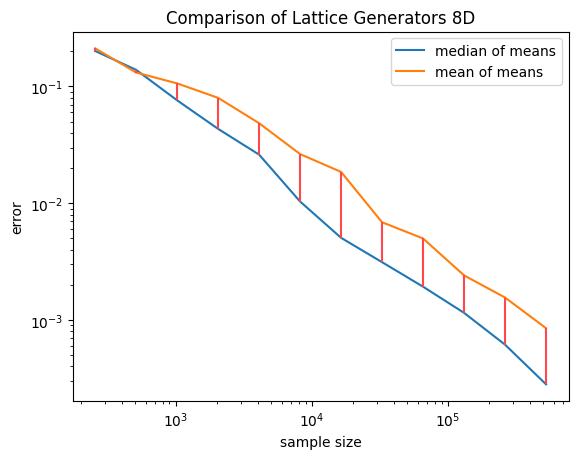}
    \caption{Comparison of lattice generators in dimensions 2,3,5,8. Results can be conditionally reproduced.}
    \label{fig:latcomp}
\end{figure}

\begin{figure}[htbp]
    \centering
    \includegraphics[width=0.45\linewidth]{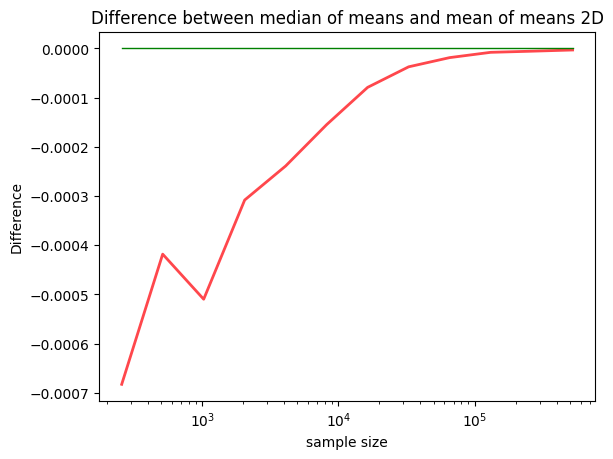}
    \includegraphics[width=0.45\linewidth]{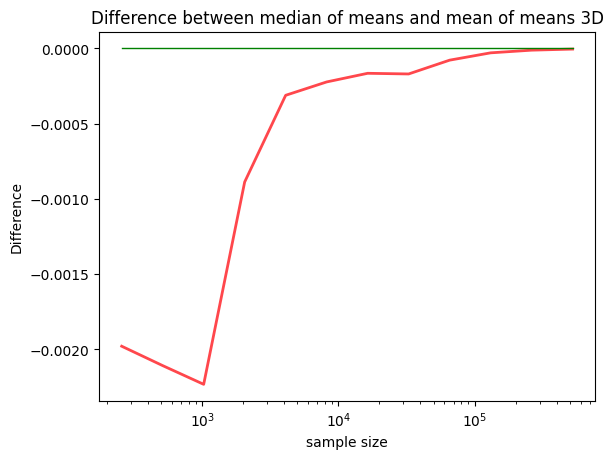}
    \includegraphics[width=0.45\linewidth]{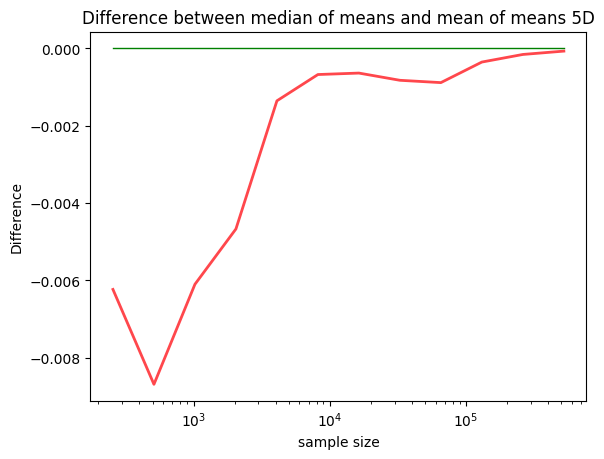}
    \includegraphics[width=0.45\linewidth]{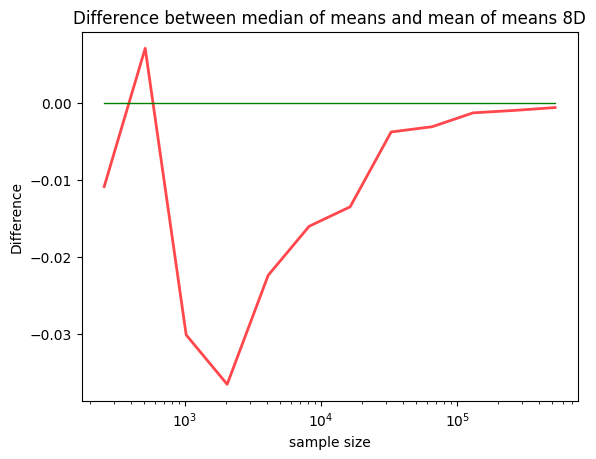}
    \caption{Difference graph of lattice generators in dimensions 2,3,5,8. Results can be conditionally reproduced.}
    \label{fig:latdiff}
\end{figure}

Results for lattices are quite similar to digital nets. Figure \ref{fig:latcomp} compares the errors produced via median-of-means and mean-of-means sampling methods. The median-of-means sampling method is colored in blue, and the mean-of-means method in orange (also reflected on the label). The vertical red lines reflect the differences between errors at a given sample size. Dimensions 2,3,5,8 were featured in the process. The graph itself is piecewise-linear (lines connecting 12 data points from 12 sample sizes ranging from $2^8$ to $2^{19}$). 

Different from digital nets, the median-of-means sampling method is almost always more accurate than the mean-of-means for dimensions 2,3,5,8 (figure \ref{fig:latdiff}). This matches the predictions in \cite{GodaLEcuyer}, that the former tends to perform the latter for large sample sizes. The only exception is that for dimension 8, the mean-of-means sampling method was superior for a small sample size ($N = 2^9 = 512$ points). This could be one of many reasons: the sample size was too small for 8 dimensions, natural sampling variation, or that mean-of-means tend to perform better for smaller sample sizes. This phenomenon, partly, is also caused by the limited accuracy in higher dimensions, due to the limited sample size.

\section{Implications, Limitations, Further Discussion}

\subsection{Implications}

This study demonstrated with statistical significance that for dimensions 2,3,5,8, the median-of-means sampling method outperforms the mean-of-means sampling method for sample sizes larger than $10^3=1000$ and smaller than or equal to $2^{19} = 524,288$, for totally randomized digital nets and lattices when integrating the Keister function. For sample sizes between $2^8 = 256$ and $1000$, the mean-of-means sampling method tends to provide higher accuracy for digital nets, and for lattices in 8 dimensions. 

This numerical experiment echos the findings of \cite{OwenPan1dmom},\\
\cite{multidMomOwenPan} and \cite{GodaLEcuyer}, that the median-of-means sampling method is more accurate than the mean-of-means sampling method for totally randomized lattices and digital nets when the sample size is sufficiently large. The behavior of the Keister function reflects these predictions quite accurately for the dimensions tested. 

This experiment also further reflects the fundamental trend in numerical integration that accuracy decreases for higher dimensions on a limited sample size \cite{Caflisch_1998}. As elaborated in Section 4.1, $2^{19}$ samples achieved an accuracy of around $10^{-6}$ for 2 dimensions, but only $10^{-3}$ for 8 dimensions. 

\subsection{Limitations}

The fundamental limitation to this study is that only the Keister function was tested. More test functions could be featured to further demonstrate the efficiency of the median-of-means, such as functions from finance--the Asian options in finance \cite{AsianOption} and the box integral in physics \cite{Bailey2007}.

Several limitations arise regarding the parameters presented in section 3.5. Only dimensions $D=2,3,5,8$ were featured in the study, while many other numerical experiments feature test integrals with higher dimensions (10,20,80,200...) \cite{doi:10.1137/S1064827599356638,GodaLEcuyer}. Similarly, the sample size $N \in [2^8,2^{19}]$ was quite limited. Some experiments feature sample sizes much larger to achieve a higher accuracy \cite{GodaLEcuyer}. Additionally, parameters $R$ and $T$, denoting the number of approximations generated for each trial, and the total number of trials ran to achieve the final result, could be modified to test the robustness of the algorithm. 

Another limitation arises due to the comparison scheme in the numerical experiment--performance was tested based on accuracy in proportion to the sample size. Another comparison could be done in parallel measuring the time required to achieve the same degree of accuracy for both algorithms \cite{doi:10.1137/S1064827599356638}. 

\subsection{Further Discussion}

To address the aforementioned limitations, future studies are recommended to compare different types of integrals (Box, Asian Option) using median-of-means versus mean-of-means, as well as other sampling methods such as traditional Monte-Carlo \cite{doi:10.1137/S1064827599356638}.
These experiments can incorporate larger sample sizes and higher dimensions, and perform time-based testing in addition to sample-size based testing. Further research is also suggested to investigate for what sample size does median-of-means outperforms mean-of-means, and vice versa.

\newpage

\begin{center}
    \textbf{References}
\end{center}

\begingroup
\renewcommand{\section}[2]{}%
\bibliographystyle{apalike}
\bibliography{./references.bib}
\endgroup

\end{document}